\documentclass[a4paper]{article}
\pdfoutput=1
\usepackage{amsmath}
\usepackage{amssymb}
\usepackage[utf8]{inputenc}
\usepackage{graphicx}
\usepackage{url}
\usepackage{doi}
\usepackage{booktabs}
\usepackage{multirow}
\usepackage{siunitx}

\usepackage{fancyhdr}
\begin{document}

\title{RoboJam: A Musical Mixture Density Network for Collaborative
  Touchscreen Interaction}
\author{Charles P. Martin\footnote{Department of Informatics,
    University of Oslo} ~and Jim Torresen\footnote{Department of
    Informatics, University of Oslo}}

\maketitle

\begin{abstract}
  
  RoboJam is a machine-learning system for generating music that assists users of a touchscreen music app by performing responses to their short improvisations. This system uses a recurrent artificial neural network to generate sequences of touchscreen interactions and absolute timings, rather than high-level musical notes. To accomplish this, RoboJam's network uses a mixture density layer to predict appropriate touch interaction locations in space and time. In this paper, we describe the design and implementation of RoboJam's network and how it has been integrated into a touchscreen music app. A preliminary evaluation analyses the system in terms of training, musical generation and user interaction.

  \end{abstract}

\section{Introduction}

\begin{figure}
\centering
\includegraphics[width=1.0\columnwidth]{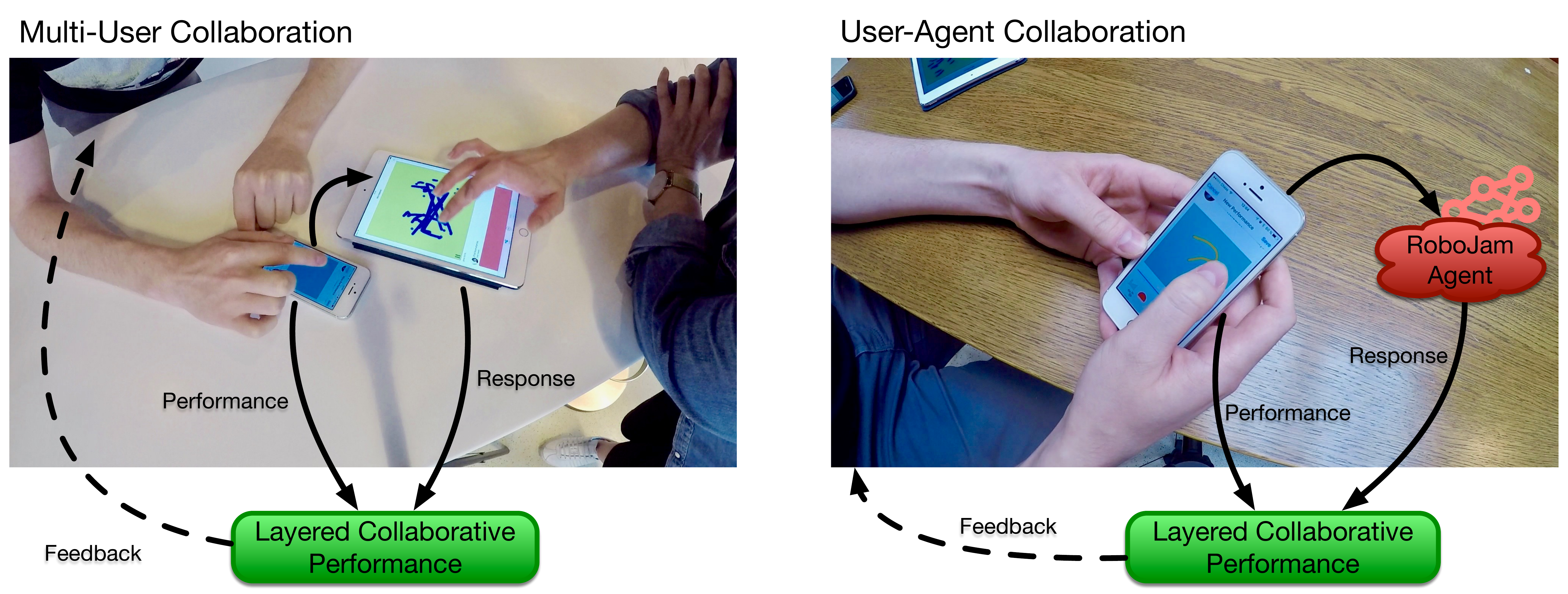}
\caption{RoboJam takes the role of a remote collaborator, providing
  musical responses to a user's touchscreen improvisations on demand.
  In this way, the user can gain feedback from a simulated
  collaboration, while waiting for other users of to respond.}
  \label{fig:computer-assisted-collaboration}
\end{figure}

New interfaces for musical expression (NIMEs) are often aimed at
casual or novice musicians, this is especially the case for
touchscreen instruments that can be deployed on popular mobile
devices. While these interfaces often emphasise a solo musical
production paradigm, the concept of ensemble, or collaborative,
performance has often been under-explored. That mobile devices are
typically used while alone, in transit, or in public, significantly
limits opportunities for touchscreen musicians to jam with others.
This limitation restricts the user's ability to gain feedback, respond
to others' ideas, and to refine their own music-making through
collaboration. In this research, we propose a machine-learning system
for generating music that can be used to assist the user by composing
responses in near real-time, thus emulating the experience of
collaborating with another user.

Our system, called \emph{RoboJam}, has been integrated into an
existing touchscreen music app. In this app, the user is able to
compose short musical pieces using the touchscreen, share them with
friends, or collaborate by ``replying'' to other performances. The
user taps, swipes and swirls in a free-form manner on the touchscreen
and these interactions are translated into musical sounds. Several
sound schemes are possible, with the touches mapped to different kinds
of synth and instrumental sounds. Recordings, which are limited to 5
seconds in length, are uploaded automatically to a cloud server where
they can be played back by other users. Users can collaborate by
replying to each others' performances, forming more complex pieces
where replies are composed as sonic layers with the original
performance. In this way, the collaborative interaction mirrors a
call-and-response style of performance; the first performer plays the
call, and subsequent performers reply with responses.

This process of musical collaboration can potentially offer important
feedback to the original performer. Hearing their own performances
layered with other responses might change how they approach further
improvisations and they might gain new ideas from others' responses.
In a co-located situation, as shown in the left hand side of Figure
\ref{fig:computer-assisted-collaboration}, this feedback loop could be
tight and allow rapid improvement and increased engagement with the
touchscreen interface. RoboJam is an agent designed to emulate this
call-and-response interaction in situation when other users aren't
present or available, this process is shown in the right side of
Figure \ref{fig:computer-assisted-collaboration}. The user can call on
RoboJam to provide a response to their performances whenever--and
however many times--it is required. RoboJam is designed to predict
what a touchscreen performer might do next, having heard the first
performance. The advantage of this design is that the user can hear
their own performance with multiple accompanying replies, thus the
user can practice and refine their own performance ideas in context.

The machine-learning system behind RoboJam uses an artificial neural
network (ANN) to generate responses on demand. Similar to other music
generation systems, RoboJam uses a recurrent neural network (RNN) with
long short-term memory (LSTM) cells to predict a temporal sequence of
discrete events. In this sequence-predicting configuration, the input
to the network is the current event, and the output is the next
predicted event; thus, the network can predict a sequence one event at
a time.

When used to predict responses, our RoboJam network first ``listens''
to the user's performance, i.e., the performance is propagated through
the network (ignoring predictions) to condition the LSTM memory state.
A new performance, 5 seconds in length, is then predicted by the
network to form the response. A key point of difference for RoboJam is
that it models the input data, a stream of touch-screen events, rather
than the musical data, frequently MIDI-note pitches, modelled by other
ANNs for generating music. This is made possible by the novel
application of a mixture density network (MDN) to creative touchscreen
interactions in RoboJam. In this research we show how this system
models musical control data, and discuss evaluations from the
perspectives of model validation, performance generation and user
experience. The results support RoboJam's ability to generate
responses that are related to the call performance and improve their
sound as well as to enhance the human performer's experience.

The structure of this paper is as follows: In Section
\ref{sec:related-work} we will discuss related work in the use of ANNs
to generate music and in co-creative musical interfaces. In Section
\ref{sec:neural-network} we will discuss the neural network design and
training for RoboJam. Section \ref{sec:interaction-design} will describe how
this network is integrated into the touchscreen app and its
interaction design. Evaluations will be discussed in Section
\ref{sec:evaluation}.


\section{Related Work}\label{sec:related-work}

\subsection{RNN Music Generation}\label{sec:RNN-music-generation}

ANNs have long been used to model and generate music. RNNs, able to
learn temporal information in between computations, are particularly
applicable to modelling musical sequences, where upcoming events
strongly depend on those that have previously occurred. In general,
these networks generate musical event sequences in a one-by-one
manner; the input is the present note, and the output predicts the next
note in the sequence. Mozer's CONCERT system~\cite{Mozer:1994aa} was
an early attempt to compose music using an RNN; this work emphasised
the advantage in learning long range dependencies in music without
handling extremely large transition tables in, for example, a
Markov model. Eck and Schmidhuber contributed further work in this
area, using an RNN to generate (potentially) endless blues
music~\cite{Eck:2007rw}. This project notably used long
short-term memory (LSTM) units \cite{Hochreiter:1997bs} to help
alleviate the problem of vanishing or exploding gradients when
training RNNs.

In recent times, the use of GPU computation and large datasets have
enabled a variety of creative applications for RNNs including the
popular CharRNN model~\cite{Karpathy:2015aa} as well as in music
generation. Sturm et al. focused on a textual representation of music
in their FolkRNN project \cite{Sturm:2016rz}. This system was trained
on a large dataset of folk melodies available online in the plain-text
``ABC'' format. Hadjeres et al. created an RNN generator of J.S.
Bach-styled chorales that can be steered towards particular
melodies~\cite{Hadjeres:2017aa}. Colombo et al. used an RNN to predict
pitch and rhythm separately~\cite{Colombo:2017aa}. Malik and Ek used
an RNN trained on MIDI-recordings of piano performances to augment
existing MIDI compositions with dynamic (volume)
instructions~\cite{Malik:2017aa}. Hutchings and McCormack used an RNN
to generate harmonic sequences in an agent-based jazz improvisation
system~\cite{Hutchings:2017aa}. Google's Magenta
project\footnote{\url{https://magenta.tensorflow.org}} have released
various trained RNN models for music generation and made steps towards
integrating them with popular music production software.

\subsection{Mixture Density Networks}\label{sec:mdn}

A commonality of the above models is that they predict sequences of
events from a finite set of symbols (e.g., one of the 128 integer MIDI
pitches). Mapping the output of an RNN to a prediction from a finite
number of classes can be effectively managed using the softmax
function. In our application, we wish to predict locations and timings
of interactions with a touchscreen, which have real---not
finite--values, so a softmax output cannot be used without a
significant loss of precision. A similar task, generating simple line
drawings, was recently tackled by the Magenta group resulting in
SketchRNN~\cite{Ha:2017ab}. This system is able to generate drawings
which, unlike pixel representations of images, are constructed from
sequences of pen movements in a 2D plane. The approach used here, and
in previous experiments in generating handwriting~\cite{Graves:2013aa}
was to replace the categorical softmax model at the outputs of the
network with a mixture model, which provide more flexible predictions.

Mixture density networks (MDNs), where the output of a neural network
is used as the parameters of a mixture model, was first explored by
Bishop as a way of learning to predict multimodal
problems that do not fit a normal distribution and are poorly modelled
by a least-squares approach to training the output of an
ANN~\cite{Bishop:1994aa}. The probability density function (PDF) of a
mixture of normal distributions can be calculated by taking the linear
combination of a number of normal distributions with weighting
coefficients. Bishop observed that an ANN could be trained to produce
these coefficients ($\pi_i$), and the centres ($\mu_i$) and variances
($\sigma_i$) for each component of the mixture. Thus a loss function
for the network could be given by the negative logarithm of the
following likelihood function, given $m$ mixtures, of target
$\mathbf{t}$ occurring in the distribution generated by input data $\mathbf{x}$:
\begin{equation}
\mathcal{L} =
\sum_{i=1}^m\pi_i(\mathbf{x})\mathcal{N}\bigl(\mu_i(\mathbf{x}),
\sigma_i^2(\mathbf{x}); \mathbf{t} \bigr)
\end{equation}
After training, the mixture model can then be sampled to generate
output data. While Bishop used an MDN with a non-recurrent network, it
has also been applied with RNNs as in Graves'
experiments~\cite{Graves:2013aa}, and SketchRNN~\cite{Ha:2017ab}
mentioned above. Such a configuration, where mixture density layers
follow an RNN, could be termed an MDRNN. While MDNs have advantages,
they also suffer from difficulties in training~\cite{Brando:2017aa},
and are not widely used. To our knowledge, an MDN design has not
previously been applied to a musical task.

\subsection{Predictive Musical Interfaces}\label{sec:predictive-musical-interfaces}

There are many examples of generative music systems that are designed to
collaborate with human performers. These are often used in a
soloist-ensemble configuration where a human performer is
``accompanied'' by artificially generated musicians.
GenJam~\cite{Biles:2007aa} is a generative backing band that knows how
to play jazz standards. The Reflexive Looper~\cite{Marchini:2017aa}
can arrange short recordings of the performer's own sounds into parts
fitting a given song structure.

Systems that \emph{learn} some aspect of the user's style and can
interact with them in performance are more relevant to this work. The
\emph{Continuator}~\cite{Pachet:2003wd} did this very effectively
using Markov models, and defined an interaction model of continuing to
play when the user stops. More recently, it has become possible to
apply RNN music models in a real-time system as shown in Magenta AI
Duet~\cite{Mann:2016aa}. In this instance, the model does not learn
from the user in the sense of updating ANN weights; rather, the user's
performance can be used to condition the LSTM cells' memory state
which then governs the style of generated notes. Due to this
conditioning and generation process, AI Duet can be directed towards
different styles after having learned a more general model of music.

\section{Neural Network Design}\label{sec:neural-network}

RoboJam's artificial neural network is a model of musical touchscreen
interactions. These interactions consist of touch points in a 2D plane
as well as the location of these interactions in time. In our
touchscreen music app, the user interacts with an area of the
touchscreen in a free-form manner to perform music, and is not
constrained by UI elements such as virtual faders or buttons. This
means that touchscreen interaction data can be considered as a record
of the user's control gestures: interactions with the touchscreen that
produce a musical result~\cite{Jensenius:2010ad}. Thus, by modelling
and generating this control data, RoboJam can produce new
performances.

The focus on control gesture data and temporal locations means that
this model differs from most music generation RNNs.
FolkRNN~\cite{Sturm:2016rz} (among others) predicts symbols from a
finite dictionary defined before training. This means that the output
of the network can be generated by sampling from a categorical
distribution given by a softmax layer. As our network must predict
real-valued locations in 2D space, a softmax layer cannot be used
without quantising the output to an unacceptably low resolution. Other
musical generation systems such as DeepBach~\cite{Hadjeres:2017aa}
generate notes on a predefined semiquaver pulse. Our network predicts
temporal locations for each touch interaction, again as a positive,
real-valued number of seconds. Generating real-valued control data
provides precision, but has a cost in terms of the amount of
high-level information (e.g., harmony) that can be learned. This
approach would not be appropriate in all musical systems, but allows
RoboJam to focus on free-form touch expression and means that it can
be used with many different musical mappings, and even in different
creative apps.

\subsection{Dataset}\label{sec:training-data}

Training data for the network consisted of musical touchscreen
improvisations. This data was derived from a corpus of performances by
touchscreen ensembles~\cite{metatone_analysis_data}. The dataset
included 163 collaborative sessions corresponding to 20 hours of
performance and 4.3M individual touch interaction events--the musical
control data for these performances. It should be noted that this
dataset was collected in a variety of apps. Each of these apps mapped
free-form touchscreen interactions directly to sound so this corpus is
an appropriate analogue for the host app for RoboJam.

The dataset was prepared first by extracting each performer's
contribution from each session. The touch locations of each record
were transformed from the original recording resolution to be in
$[0,1] \times [0,1]$. Absolute times were changed to time deltas with
a maximum value of 5 seconds. This means that each touch event became
a vector $(x, y, dt) \in [0,1] \times [0,1] \times [0,5]$. For
training, the performances were placed end-to-end to form one long
sequence of touch events. Overlapping examples of 256 events in length
were extracted from this sequence resulting in almost 4.3M training
examples.

\subsection{Implementation}

\begin{figure}
  \centering
  \includegraphics[width=1.0\columnwidth]{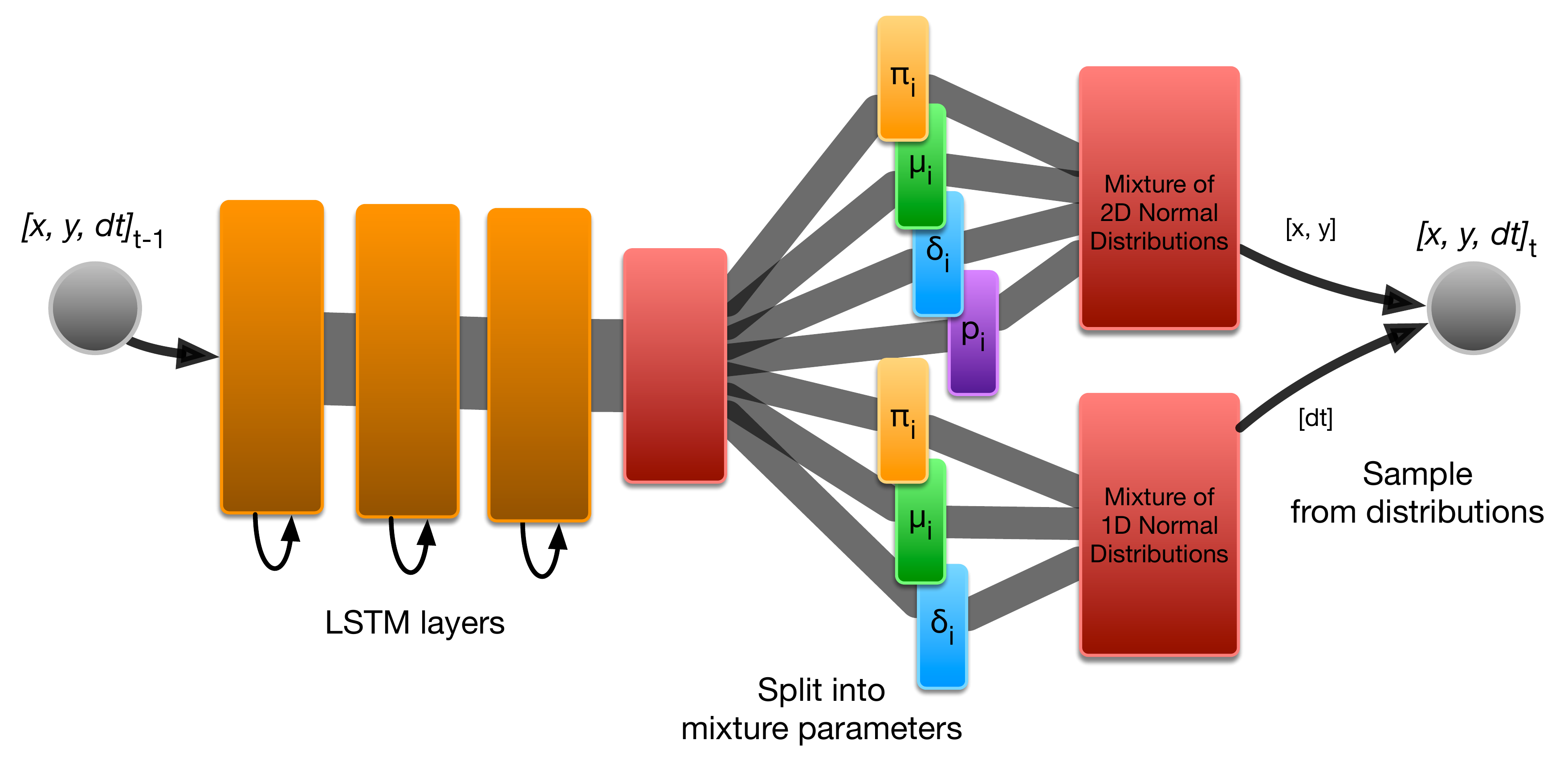}
  \caption{RoboJam uses the output of an RNN to parameterise mixture
    models that handle touchscreen location and the timing of events.
    Two models are used: a mixture of 2D normal distributions to
    predict location of touches, and a mixture of 1D normals to
    predict time deltas in between events.}
  \label{fig:mdrnn-design}
\end{figure}

To model this data, we use an MDRNN design, inspired by
SketchRNN~\cite{Ha:2017ab} and Graves' handwriting generation
network~\cite{Graves:2013aa} (discussed in Section \ref{sec:mdn}).
Both of these systems are intended to generate sequences of 2D surface
interactions, very similar to our touchscreen data; however, sketches
and handwriting do not have a significant temporal component except
for the ordering of strokes. Both of these previous ANNs used the
outputs of LSTM cells as the parameters of a mixture of probability
models. In both cases, a mixture of bi-dimensional normal
distributions were used to predict the location of the next pen
movement in 2D space, the number of normal distributions in a mixture
was a hyperparameter in training these networks. In our case, location
of touches is accompanied by a positive time value, indicating the
number of seconds in the future that the interaction should occur. The
same 2D mixture distribution as in SketchRNN is used to
predict spatial location of touches while a separate mixture of 1D
normal distributions is used to predict the temporal location of the
touch.

Our MDRNN is implemented in TensorFlow, and the source code is
available online\footnote{RoboJam's source code is available at \url{https://doi.org/10.5281/zenodo.1064014}}. An
overview of our MDRNN design is shown in Figure
\ref{fig:mdrnn-design}. Input data are vectors
$[x,y,dt] \in \mathbb{R}^3$ corresponding to absolute location in the
plane $[0,1] \times [0,1]$ and value in $[0,5]$ indicating the number
of seconds since the last interaction. The input  vector flows through
three fully-connected layers of LSTM cells. The output of the last
LSTM layer is projected using a fully connected layer to a vector $p$
that is used to generate parameters for our two mixture models that
predict time and space values respectively.

The time model is a mixture of $M$ 1D normal distributions,
parameterised as given in Section \ref{sec:mdn} by a mean ($\mu_t$),
standard deviation ($\sigma_t$), and a coefficient $\pi_t$ for each mixture
component. The space model is a mixture of $M$ 2D normal
distributions, each having two means ($\mu_x, \mu_y$), standard deviations
($\sigma_x, \sigma_y$), one correlation ($\rho$) and coefficient
($\pi_{xy}$). So, $p$ has size $3M + 6M$.

The components of $p$ are transformed as recommended by
Brando~\cite{Brando:2017aa}, Graves~\cite{Graves:2013aa} and
Bishop~\cite{Bishop:1994aa}. This ensures that the standard deviations are
greater than zero, correlations are in $(-1,1)$, and that the coefficients
for each mixture sum to one. In operation, a sample is drawn from
these two mixture models to generate the next vector: 
$[x_{t+1}, y_{t+1}, dt_{t+1}]$. In training, we measure the
likelihood of the next (known) vector occurring in the generated
mixture models. The loss function has two components that are added
--- the average negative log likelihood over the batch for the time model
and for the space model:
\begin{equation}
  \mathcal{L}_{space} = -\frac{1}{N} \sum_{i=1}^N \log \sum_{j=1}^{M}
  \bigl(\pi_{xy,j}\mathcal{N}_{2D}(\mu_{x,j},\mu_{y,j},\sigma_{x,j},\sigma_{y,j},\rho_{j}; x_i,y_i)\bigr)
\end{equation}
\begin{equation}
  \mathcal{L}_{time} = -\frac{1}{N} \sum_{i=1}^N \log
\sum_{j=1}^M\pi_{dt,j}\mathcal{N}_{1D}(\mu_{dt,j},
\sigma_{dt,j};  dt_i)
\end{equation}
\begin{equation}
\mathcal{L}_{tot} =  \mathcal{L}_{space} + \mathcal{L}_{time}
\end{equation}
Where $\mathcal{N}_{2D}$ is as given in equations 24 and 25 of
\cite{Graves:2013aa}, and $\mathcal{N}_{1D}$ is as given in equation
23 of \cite{Bishop:1994aa}. Code for $\mathcal{N}_{2D}$ follows Ha and
Eck's work~\cite{Ha:2017ab} and a similar TensorFlow implementation
was developed for $\mathcal{N}_{1D}$.

\subsection{Training}\label{sec:training}

As discussed above in Section \ref{sec:training-data}, the training
dataset consisted of 4.3M overlapping examples of 256-event
performance excerpts. These examples were shuffled before training and
divided into 33579 batches of 128 examples. The Adam optimiser was
used for gradient descent with an initial learning rate of
$1 \times 10^{-4}$. RoboJam's network was configured with 3 layers of
LSTM units and $M = 16$ mixtures. Models were trained up to five
epochs with LSTM layer sizes of 64, 128, 256, and 512 units. The
results of this training is discussed below in Section
\ref{sec:validation}.

Numerical stability in training is a significant challenge for MDNs.
We follow advice from Bishop~\cite{Bishop:1994aa},
Brando~\cite{Brando:2017aa} and others to transform and clip the
mixture parameters, and to apply gradient clipping. Nevertheless,
avoiding division by zero when calculating $\mathcal{L}_{tot}$
continued to be a challenge in this research and is a topic for future
investigation.

\subsection{Model Limitations}\label{sec:limitations}

Our implementation separates space and time predictions into two
mixture models, both connected to the output of one RNN. It may be
possible to represent these in a single mixture of 3D normal
distributions, but the PDF for a 3D normal is more complex. Our
decision here was pragmatic in that we were able to extend previously
effective MDN designs with parameters for space and time predictions
linked via the fully-connected LSTM and projection layers. Future work
could determine whether a 3D model, or indeed three 1D models would be
more suitable. Our model does not explicitly consider the state of a
touch event (whether it is a new, or moving touch), even though this
data is available in the corpus. Given the temporal dimension, this
aspect of the data is quite predictable so we used a heuristic
threshold of $dt > 0.1$ to determine a touch as new. Potentially this
data could be predicted by the MDRNN as in earlier
handwriting~\cite{Graves:2013aa} and sketch~\cite{Ha:2017ab} models.


\section{Interaction Design}\label{sec:interaction-design}

\begin{figure}
\centering
\includegraphics[width=1.0\columnwidth]{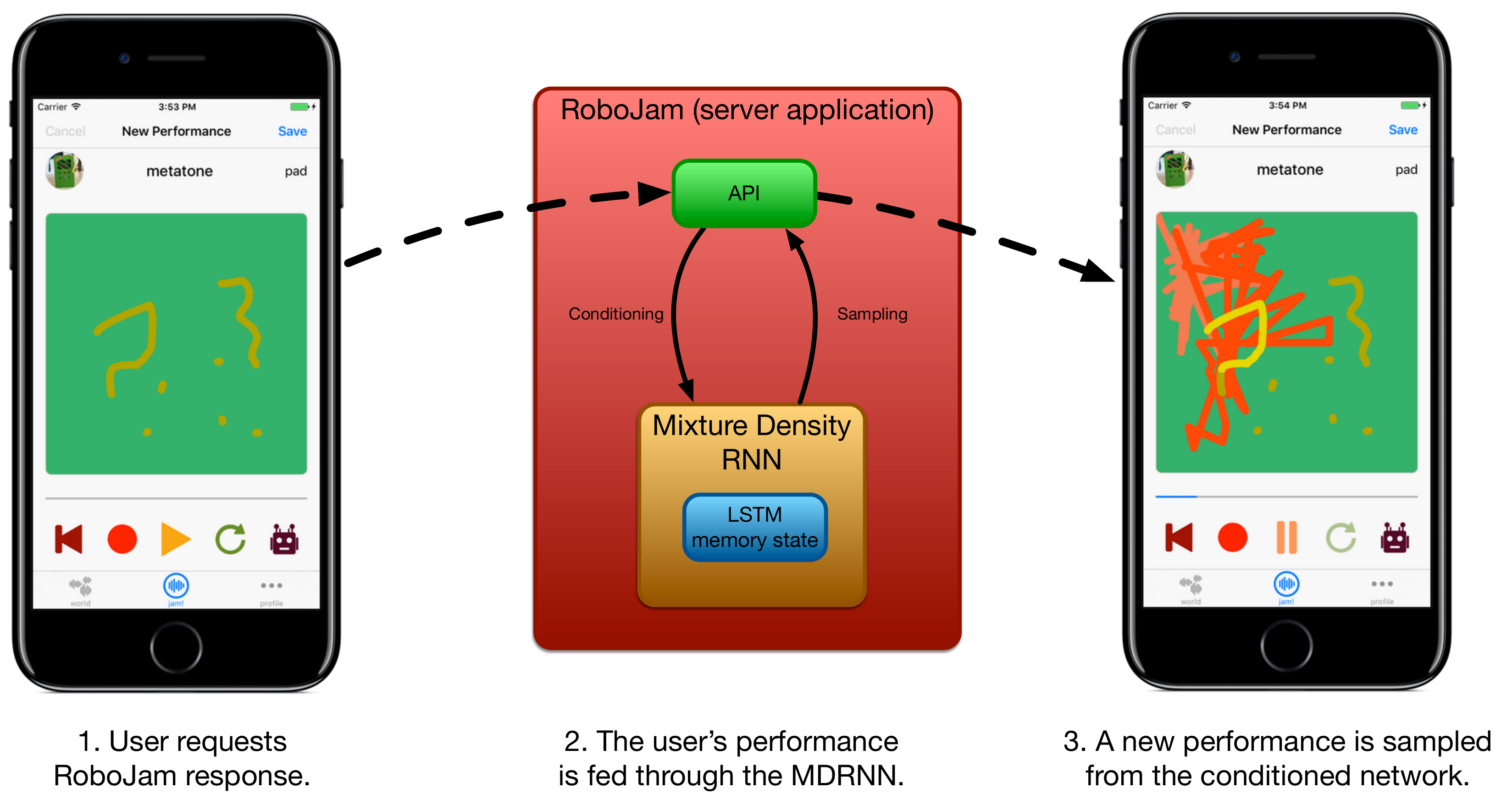}
\caption{RoboJam is a web-application that generates responses to
  touchscreen performances. User performances are sent to the server,
  used to condition the memory state of an MDRNN, this can then
  generate a new performance that is returned to the user as layer of
  accompaniment.}
\label{fig:robojam-interaction}
\end{figure}

RoboJam is a web-application that uses the MDRNN described above to
predict appropriate responses to user-created touchscreen
performances. The interaction paradigm mirrors a call-and-response
improvisation; the performer creates a short improvisation, RoboJam
``listens'' to this performance and creates a response, then both
contributions can be played back at the same time\footnote{This
  process is illustrated in the video figure:
  \url{https://vimeo.com/242251501}}.

The system architecture is illustrated in Figure
\ref{fig:robojam-interaction}: First, the user performs a short piece
of music with the touchscreen interface; the app limits these
performances to 5 seconds in length. The user can then request a
RoboJam response by tapping a button in the app's graphical interface.
The performance, consisting of a sequence of touch interaction events,
is encoded in JSON format and sent to the server via a REST API. When
the server receives a performance, it feeds each touch event in
sequence through the MDRNN in order to condition the neural net's
memory state. When this is complete, it generates a new performance
from the MDRNN by sampling a new sequence of touch-events until 5
seconds of interactions have been saved. This sampled performance is
sent back to the user's device and displayed as a second layer of
performance on the screen. When the user hits ``play'', both
performances are played back simultaneously as separate parts.


In this research, RoboJam has been used with a prototype touchscreen
music app for iOS devices. This app allows touch interactions to
be interpreted as different instrumental sounds such as drums, strings,
bass, or different synthesiser sounds. In general, these mappings
allow the user continuous control of pitch on the x-axis and tone or
effects on the y-axis. The mappings are available to the user as selectable
presets for their performances. Response performances from RoboJam's
MDRNN can be performed by any one of these mappings, but our
implementation assigns one randomly to each response such that the
mapping is different than that currently selected by the user.

RoboJam is implemented in Python using the Flask web development
framework. This design allowed RoboJam's TensorFlow-based MDRNN (also
defined in Python) to be accessed easily. Future implementations of
RoboJam could be integrated directly into a touchscreen app using
TensorFlow, or manufacturer-specific deep learning frameworks such as
Apple's CoreML. In the short-term, however, the simplicity of using
the MDRNN in a client/server architecture outweighed the potential
benefits of on-device predictions.

\section{Evaluation}\label{sec:evaluation}

Evaluation of RoboJam, as with many co-creative interfaces, presents
challenges. We have chosen to split evaluation into three stages:
validation, generation, and interaction. These stages respectively
address the following questions: Has the networked learned anything?
Does the network generate realistic performances? Is the interactive
system useful for its purpose of enhancing the users' musical
experience?

\subsection{Validation}\label{sec:validation}

\begin{figure}
\centering
\includegraphics[width=0.9\columnwidth]{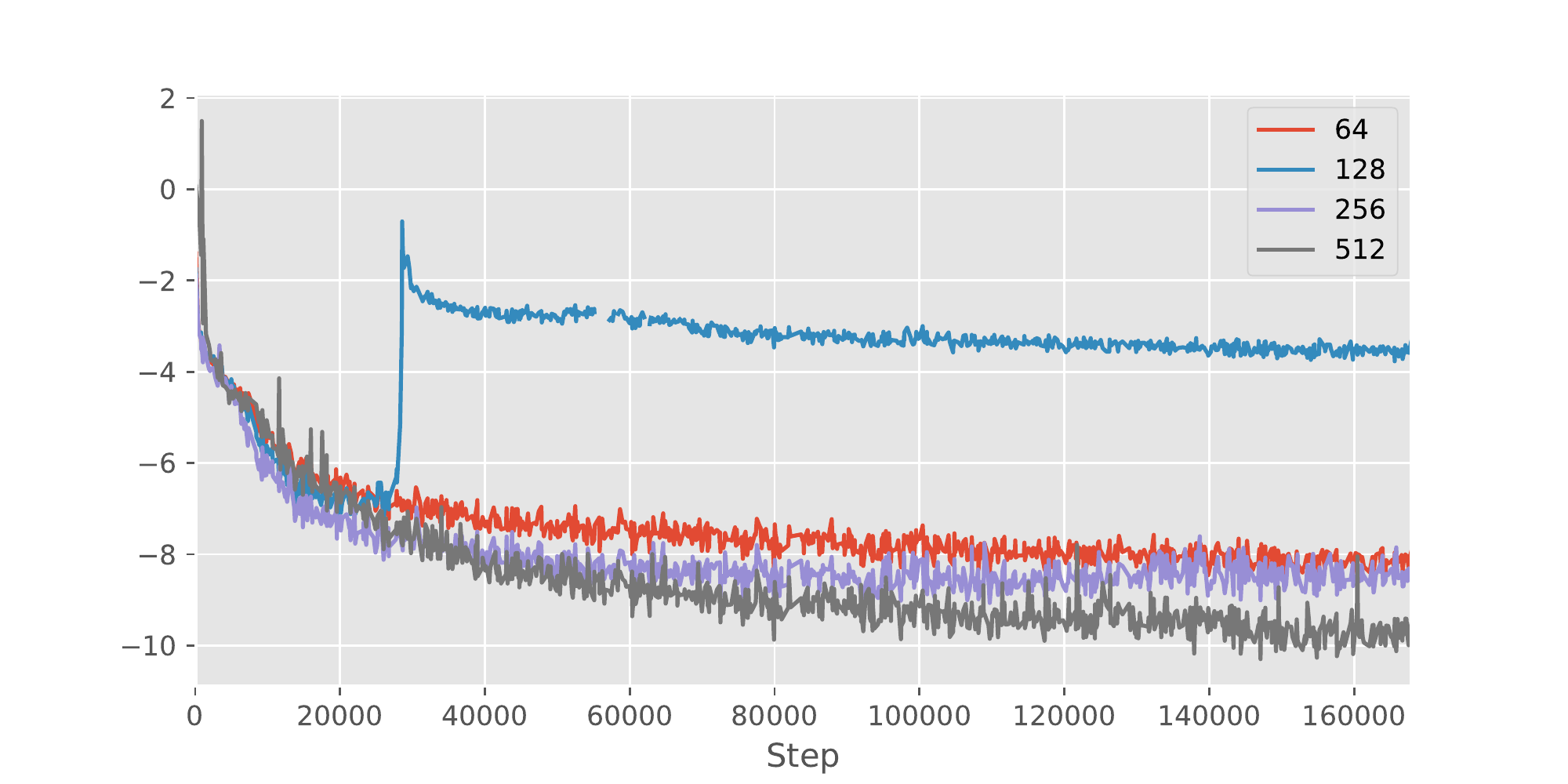}
\caption{Training loss for RoboJam networks with 64, 128, 256, and
  512-unit LSTM layers. Training and validation loss for the
  512 unit network was lowest. The 128-unit network
  failed to train due to numerical errors.}
\label{fig:training-loss}
\end{figure}

\begin{table}[]
\centering
\begin{tabular}{@{}p{1.5cm}llllll@{}}
  \toprule
  \multirow{2}{2.0cm}{Training Step (1000s)}
                      & \multicolumn{2}{c}{64 units} & \multicolumn{2}{c}{256 units} & \multicolumn{2}{c}{512 units} \\
                      & Training& Valid'n & Training& Valid'n& Training  & Valid'n  \\ \midrule
2                     & -3.72       & -4.21  & -3.98     & -4.39   & -3.56     & -3.81   \\
34                    & -6.73       & -7.58  & -7.30     & -8.58   & -6.96     & -7.92   \\
68                    & -7.61       & -8.21  & -8.42     & -8.98   & -8.86     & -9.50   \\
100                   & -7.67       & -8.39  & -8.14     & -8.85   & -9.14     & -9.88   \\
132                   & -7.93       & -8.61  & -8.19     & -9.09   & -9.47     & -10.00  \\
166                   & -8.07       & -8.76  & -8.51     & -8.84   & -9.42     & -10.31  \\ \bottomrule\\
\end{tabular}
\caption{Training and validation loss for RoboJam's MDRNN at
  throughout training and with different RNN-layer sizes. (Lower is
  better.)}
\label{tab:validation-experiment}
\end{table}



Four versions of the model were trained up to five epochs (166000
batches) using 64, 128, 256, and 512-unit LSTM layers respectively.
Training loss for these models can be seen in \ref{fig:training-loss};
notably, the 128-unit network failed to train due to numerical errors.

Validation was performed by calculating loss values when the network
predicts excerpts of touchscreen performances from RoboJam's host
touchscreen app. This validation dataset was not used in training and
consisted of 1047 performances with a total of 168039 touch
interactions; these data were processed similarly to the training set.
A validation experiment was performed on the three successfully
trained models with training and validation loss recorded at model
checkpoints near the end of each epoch. The results in Table
\ref{tab:validation-experiment} show that both training loss and
validation loss decrease during training with the lowest scores
occurring with the 512-unit network after five epochs. This model was
used for the two evaluations below.

It is notable that the loss on the validation set is consistently
lower than on the training set -- an unusual situation. This suggests
that the validation data may actually be more predictable than the
training data. Given that the training data was collected from a
variety of different apps, it may contain more unpredictable or
arbitrary interactions than we thought. While validation shows that
the network's predictive ability improved during training, it may be
better in future to focus training on a dataset that has been
restricted to the most relevant interactions.

\subsection{Performance Generation}\label{sec:generation}

\begin{figure}
\centering
\includegraphics[width=1.0\columnwidth]{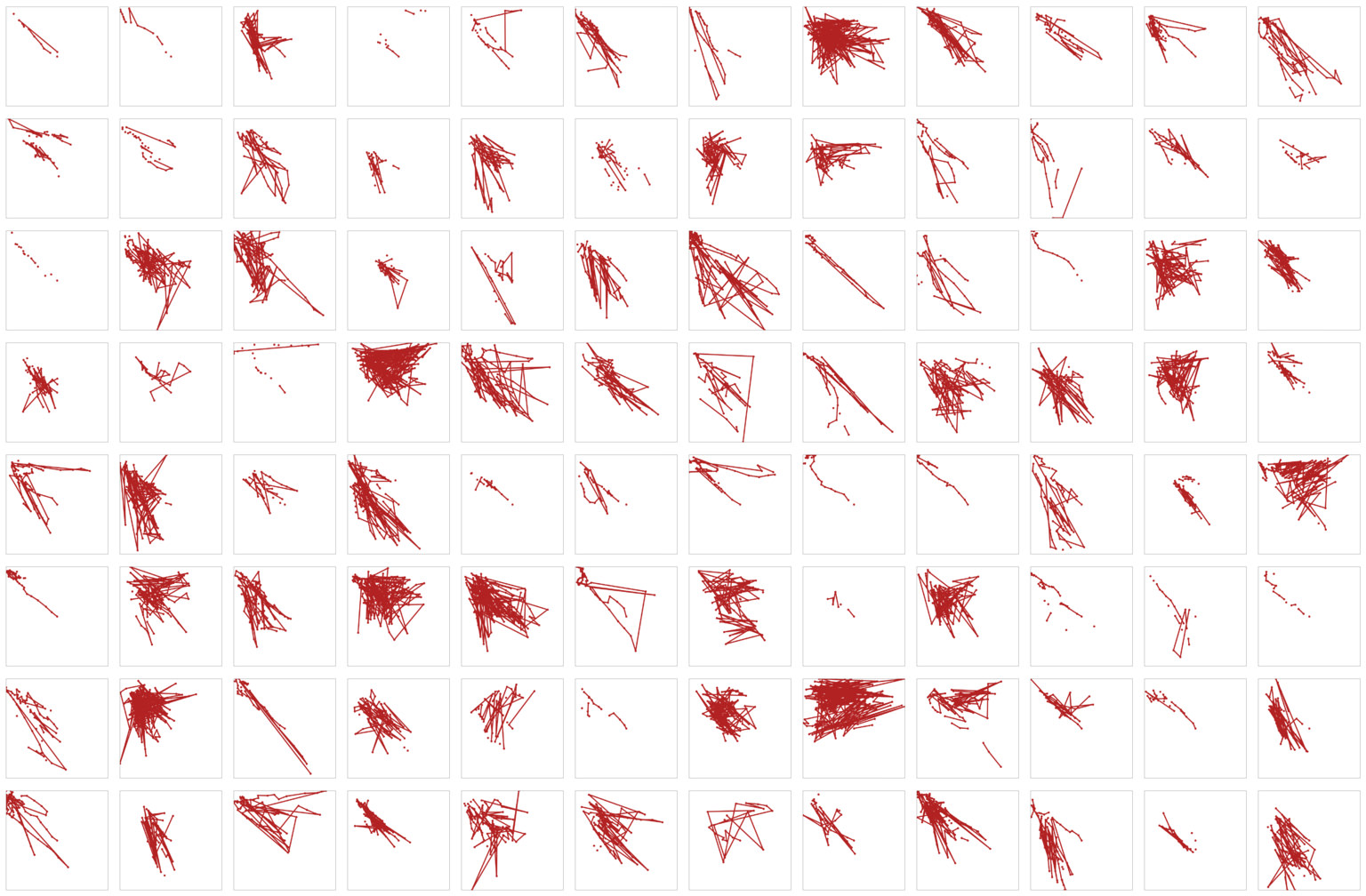}
\caption{RoboJam's network generating 5-second performances in
  ``unconditional'' mode, i.e., with an empty memory state and
  starting from the centre of the performance area.}
\label{fig:uncond-generation}
\end{figure}

\begin{figure}
\centering
\includegraphics[width=1.0\columnwidth]{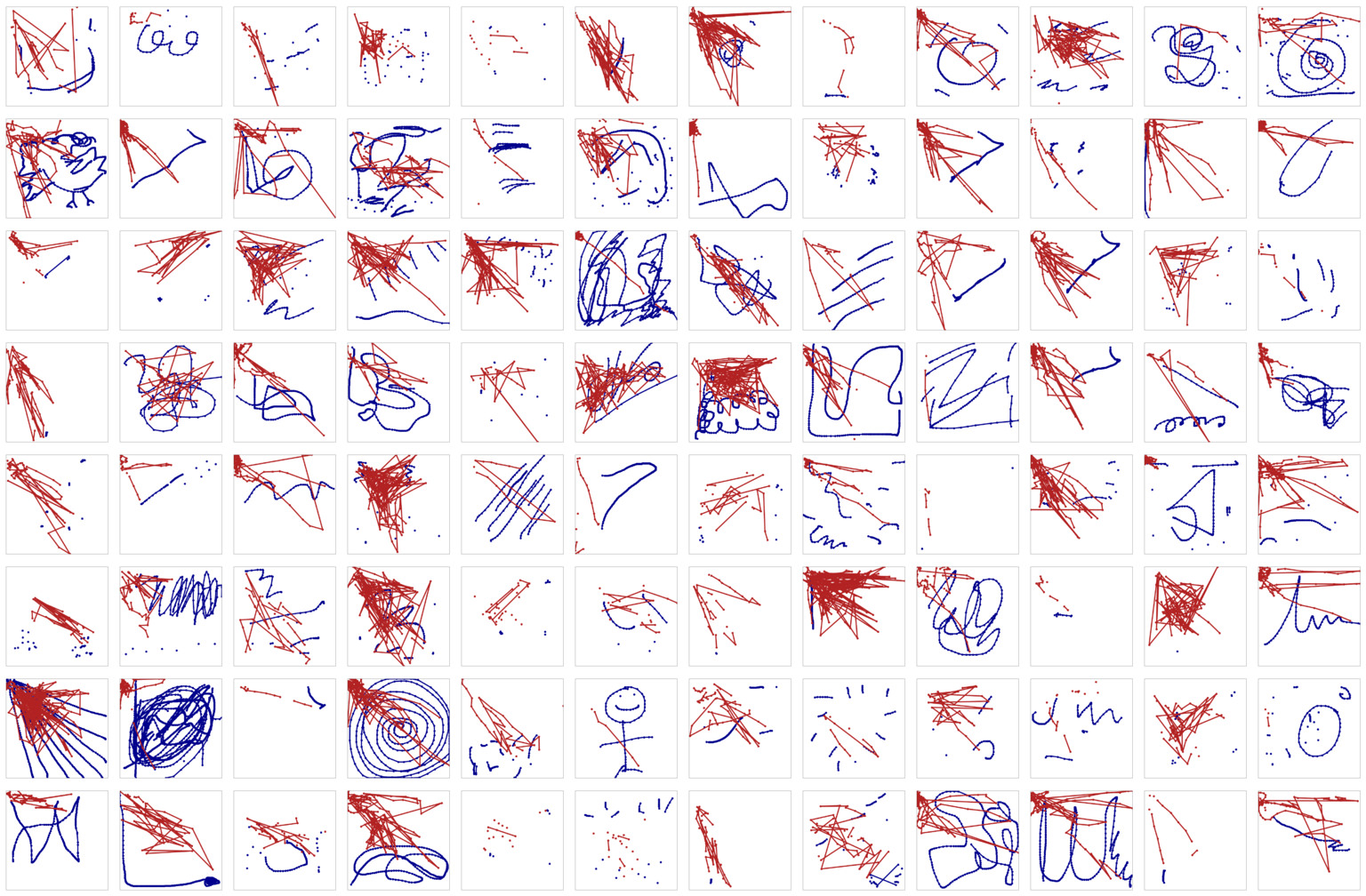}
\caption{Examples of 5-second responses to performances from the
  validation set. The input performance
  is shown in blue and RoboJam's response is shown in red.}
\label{fig:cond-generation}
\end{figure}

The output of RoboJam's MDRNN are performances that should be
experienced in time; however, as a quick overview, we can view plots
of the touches from these performance over their whole duration.
Figure \ref{fig:uncond-generation} shows five-second performances
generated without conditioning the network's memory state. More
relevant to the application, Figure \ref{fig:cond-generation} shows
examples conditioned performances, the input performance is shown in
blue and RoboJam's response is shown in red. In both figures, moving
touches are connected while individual touches are not. Examples of
call-and-response RoboJam performances can be found in a video figure
for this paper which is available online\footnote{Video figure:
  \url{https://vimeo.com/242251501}}.

Figures \ref{fig:uncond-generation} and \ref{fig:cond-generation} show
a variety of touchscreen interaction behaviours, but reveal
limitations in terms of generating realistic performances. Compared
with example touchscreen performances (shown in blue in Figure
\ref{fig:cond-generation}), the network generated performances do not
have as many smooth paths, and have many more large jumps across the
screen with very short time delays (resulting in long connected
paths). While the unconditional performance (Figure
\ref{fig:uncond-generation}) show uncontrolled behaviour, many of the
conditioned performances \ref{fig:cond-generation} bear some relation
to the ``style'' of the input performance, in terms of location of
touch points, direction and shape of motion. Some rhythmic
relationships can also be seen, input performances with sparse
disconnected paths indicate a rhythmic performance, these seem to lead
to more rhythmic variation in the generated response.

Frequently in the generated performances, the touch point tends to
move to the upper left corner of the touch area, corresponding to
location $(0,0)$ in the 2D plane. This tendency is as confusing as it
is worrying, given that the training data is much more evenly spread
across the touch field. Exploring the reasons and possible solutions
for this issue is a topic to be addressed in future research.

\subsection{Interaction}\label{sec:interaction-evaluation}

\begin{figure}
\centering
\includegraphics[width=1.0\columnwidth]{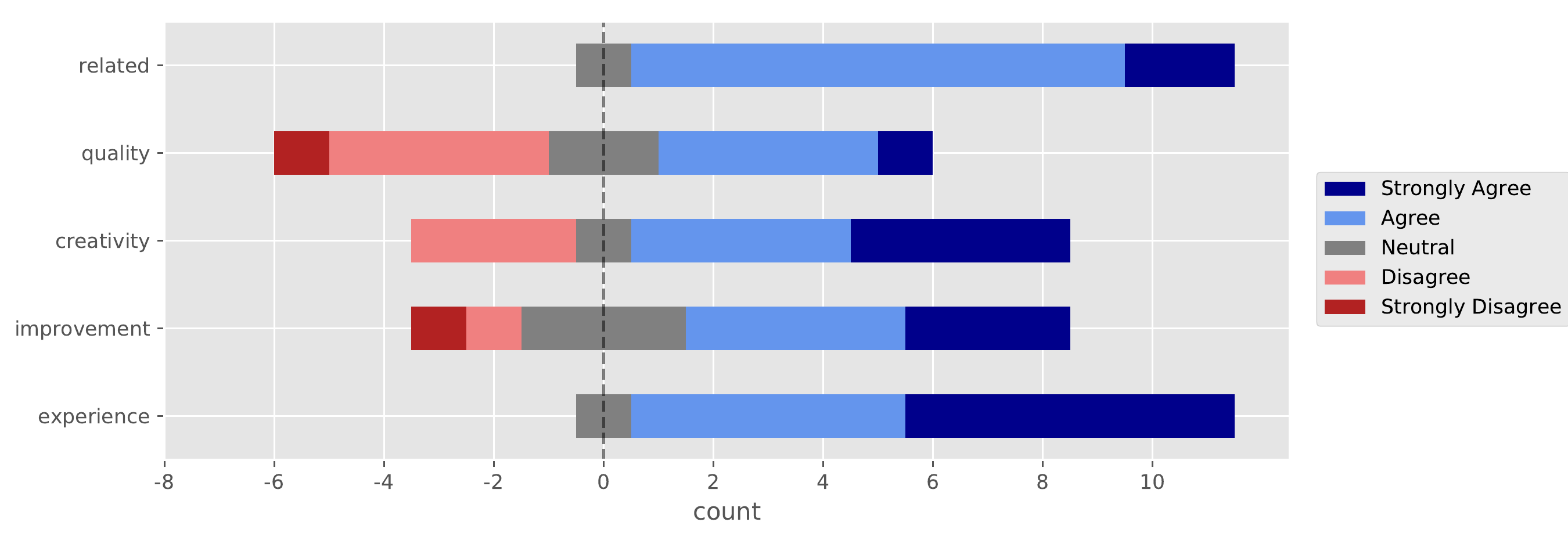}
\caption{Distribution of responses to each question in the preliminary
  user study of RoboJam. Most users agreed that the responses were
  related to their performances and that interacting with RoboJam
  enhanced their experience.}
\label{fig:likert-scale-data}
\end{figure}

A preliminary evaluation of interaction was performed with a small
group of users with RoboJam connected to a touchscreen music
application. The aim of the study was to gain a rough picture of how
users viewed the quality of RoboJam's responses and the experience of
interacting with this call-and-response agent. Twelve participants
were included in the study: students, researchers, and teachers of
computer science and music with a mix of musical experience. The study
procedure was as follows: Each participant was given a short tutorial and
practice session (around 5 minutes) in creating performances in the
app. They were then asked to create several performances and
generate as many responses from RoboJam as they liked, listening to
each resulting layered performance. After the test session, the
participants responded to five Likert-style statements on 5-point
agreement scales (\emph{strongly disagree, disagree, neutral, agree,
  strongly agree}). The questions were:
\begin{enumerate}
\item The responses were related to my performance. (\emph{related})
\item The responses had a high musical quality. (\emph{quality})
\item The responses showed musical creativity. (\emph{creativity})
\item The response layer made my performance sound better. (\emph{improvement})
\item Interacting with RoboJam enhanced my experience. (\emph{experience})
\end{enumerate}

The results of this study are shown in Figure
\ref{fig:likert-scale-data}. Almost all the users felt that the
responses were related to their performance and that interacting with
RoboJam enhanced their experience. Most participants agreed that
RoboJam's responses showed creativity and made their performances
sound better; however, they were more uncertain about the overall
quality of the responses. These results tell us that interacting with
a call-and-response agent is a potentially useful addition to this
app. In some ways, it would appear that this feature is useful even
when the agent produces responses of uneven quality. The participants
appear to have appreciated hearing their performances in context with
a response and felt that this improved the sound of their own
contributions. They were able to naturally cherry-pick responses until
they found one that appealed to them, this involved critically
engaging with their own performances as well as RoboJam's.

These results encourage us that RoboJam could be a useful and
significant addition to our touchscreen music app. A more in depth
user study could compare RoboJam responses with other agents that
generate responses via alternative generative music processes.

\section{Conclusion}\label{sec:conclusion}

This work has described RoboJam, an ANN-based musical sequence
generator connected to an interactive touchscreen music app as a agent
for call-and-response style improvisation. This system has the
potential to enhance the user's experience, particularly when they are
not able to collaborate with other users. Our agent, RoboJam, uses a
novel application of MDN and RNN to model musical control data. Not
only does this network model the location of interactions on a
touchscreen, but it also the \emph{rhythm} of these interactions in
absolute time. This configuration distinguishes RoboJam from other
typical approaches in ANN music generation; our network learns from
data at the control gesture level, rather than at the note level, it
also learns to perform in absolute time, rather than at preset
rhythmic subdivisions. In the context of an interactive digital
musical instrument, this configuration allows RoboJam to perform music
in exactly the same way as users. Rather than learning to compose
music, RoboJam is actually learning to perform on a touchscreen
instrument.

This design choice has several implications in our results. Learning
from low-level interaction data seems to be a harder task than
learning from higher-level musical notes. From a musical perspective,
RoboJam's performances seem underwhelming compared with high-level ANN
music generators. However, there are advantages from learning to
\emph{perform} rather than to \emph{compose}. Since a touchscreen
interface can be used to control different kinds of instruments,
RoboJam's output can be mapped to different kinds of synthesis
processes as in our touchscreen app. Furthermore, RoboJam's responses
are related to the body movements of the app user, an embodied
approach that is appropriate in our application where free-form
gestural exploration is emphasised.

In this research, we examined the results of evaluations of RoboJam in
terms of model validation, generative power, and user experience.
These studies have demonstrated that this system can generate
responses that are related in movements and rhythm to the call
performance. Our preliminary human-centred evaluation has shown that
users felt the responses improved their performances and enhanced
their experiences. This was the case even when response quality was
not always highly rated.

Our results are encouraging, but they are tempered by the difficulty
of training an MDRNN. Our musical data contains a wider and more
abstract variety of interactions than, for example, handwritten
letters of the alphabet. This may explain some of our difficulties
with training. Future improvements to RoboJam could explore more
curated training data and alternative mixture model designs.

The generation of music and other creative data is an exciting topic
in deep learning. While it is clear that computer-generated art spurs
the imagination, it is less clear how music generators can be
integrated into human-centred creative processes. We have focused on
modelling musical touchscreen control data and on developing a
call-and-response interaction between user and agent. As a result, our
system can easily be integrated into mobile musical interfaces and
appears to enhance the musical experience of users in our study.
Future research could expand on how models of musical control data and
call-and-response agents could be used to accompany, modulate, or
assess human musical performances.

\subsection*{Acknowledgements}

This work was supported by The Research Council of Norway as a
part of the Engineering Predictability with Embodied Cognition (EPEC)
project, under grant agreement 240862.

\bibliographystyle{plainurl}
\bibliography{biblio.bib}

\begin{thebibliography}{10}

\bibitem{Biles:2007aa}
John~A. Biles.
\newblock Improvizing with genetic algorithms: Genjam.
\newblock In Eduardo~Reck Miranda and John~Al Biles, editors, {\em Evolutionary
  Computer Music}, pages 137--169. Springer London, London, 2007.
\newblock \href {http://dx.doi.org/10.1007/978-1-84628-600-1_7}
  {\path{doi:10.1007/978-1-84628-600-1_7}}.

\bibitem{Bishop:1994aa}
Christopher~M. Bishop.
\newblock Mixture density networks.
\newblock Technical Report NCRG/97/004, Neural Computing Research Group, Aston
  University, 1994.

\bibitem{Brando:2017aa}
Axel Brando.
\newblock Mixture density networks (mdn) for distribution and uncertainty
  estimation.
\newblock Master's thesis, Universitat Polit\`{e}cnica de Catalunya, 2017.
\newblock URL:
  \url{https://github.com/axelbrando/Mixture-Density-Networks-for-distribution-and-uncertainty-estimation/}.

\bibitem{Colombo:2017aa}
Florian Colombo, Alexander Seeholzer, and Wulfram Gerstner.
\newblock Deep artificial composer: A creative neural network model for
  automated melody generation.
\newblock In João Correia, Vic Ciesielski, and Antonios Liapis, editors, {\em
  Computational Intelligence in Music, Sound, Art and Design: 6th International
  Conference, EvoMUSART 2017 Proceedings}, pages 81--96. Springer International
  Publishing, Cham, 2017.
\newblock \href {http://dx.doi.org/10.1007/978-3-319-55750-2_6}
  {\path{doi:10.1007/978-3-319-55750-2_6}}.

\bibitem{Eck:2007rw}
Douglas Eck and Jürgen Schmidhuber.
\newblock A first look at music composition using lstm recurrent neural
  networks.
\newblock Technical Report IDSIA-07-02, Instituto Dalle Molle di studi sull'
  intelligenza artificiale, Manno, Switzerland, 2007.

\bibitem{Graves:2013aa}
A.~{Graves}.
\newblock {Generating Sequences With Recurrent Neural Networks}.
\newblock {\em ArXiv e-prints}, August 2013.
\newblock URL: \url{https://arxiv.org/abs/1308.0850}.

\bibitem{Ha:2017ab}
D.~{Ha} and D.~{Eck}.
\newblock {A Neural Representation of Sketch Drawings}.
\newblock {\em ArXiv e-prints}, April 2017.
\newblock URL: \url{https://arxiv.org/abs/1704.03477}.

\bibitem{Hadjeres:2017aa}
Gaëtan Hadjeres, Fran{\c{c}}ois Pachet, and Frank Nielsen.
\newblock {D}eep{B}ach: a steerable model for {B}ach chorales generation.
\newblock In Doina Precup and Yee~Whye Teh, editors, {\em Proceedings of the
  34th International Conference on Machine Learning}, volume~70 of {\em
  Proceedings of Machine Learning Research}, pages 1362--1371, International
  Convention Centre, Sydney, Australia, 06--11 Aug 2017. PMLR.
\newblock URL: \url{http://proceedings.mlr.press/v70/hadjeres17a.html}.

\bibitem{Hochreiter:1997bs}
Sepp Hochreiter and Jürgen Schmidhuber.
\newblock Long short-term memory.
\newblock {\em Neural Computation}, 9(8):1735--1780, 1997.
\newblock \href {http://dx.doi.org/10.1162/neco.1997.9.8.1735}
  {\path{doi:10.1162/neco.1997.9.8.1735}}.

\bibitem{Hutchings:2017aa}
Patrick Hutchings and Jon McCormack.
\newblock Using autonomous agents to improvise music compositions in real-time.
\newblock In {\em EvoMUSART 2017}, volume 10198 of {\em LNCS}. Springer
  International Publishing, 2017.
\newblock \href {http://dx.doi.org/10.1007/978-3-319-55750-2_8}
  {\path{doi:10.1007/978-3-319-55750-2_8}}.

\bibitem{Jensenius:2010ad}
Alexander~Refsum Jensenius, Marcelo~M. Wanderley, Rolf~Inge Godøy, and Marc
  Leman.
\newblock Musical gestures: Concepts and methods in research.
\newblock In {\em Musical Gestures: Sound, Movement, and Meaning}. Routledge,
  2010.

\bibitem{Karpathy:2015aa}
Andrej Karpathy.
\newblock The unreasonable effectiveness of recurrent neural networks.
\newblock Published on Andrej Karpathy's blog, May 2015.
\newblock URL: \url{http://karpathy.github.io/2015/05/21/rnn-effectiveness/}.

\bibitem{Malik:2017aa}
I.~{Malik} and C.~H. {Ek}.
\newblock Neural translation of musical style.
\newblock {\em ArXiv e-prints}, August 2017.
\newblock URL: \url{https://arxiv.org/abs/1708.03535}.

\bibitem{Mann:2016aa}
Yotam Mann.
\newblock {AI} duet.
\newblock Git Repository, 2016.
\newblock URL:
  \url{https://github.com/tensorflow/magenta-demos/tree/master/ai-jam-js}.

\bibitem{Marchini:2017aa}
Marco Marchini, François Pachet, and Benoît Carré.
\newblock Rethinking reflexive looper for structured pop music.
\newblock In {\em Proceedings of the International Conference on New Interfaces
  for Musical Expression}, pages 139--144, Copenhagen, 2017. Aalborg University
  Copenhagen.
\newblock URL:
  \url{http://www.nime.org/proceedings/2017/nime2017_paper0027.pdf}.

\bibitem{metatone_analysis_data}
Charles Martin, Ben Swift, and Henry Gardner.
\newblock anucc/metatone-analysis: Touchscreen data corpus, 2017.
\newblock URL: \url{https://doi.org/10.5281/zenodo.1020166}, \href
  {http://dx.doi.org/10.5281/zenodo.1020166}
  {\path{doi:10.5281/zenodo.1020166}}.

\bibitem{Mozer:1994aa}
Michael~C. Mozer.
\newblock Neural network music composition by prediction: Exploring the
  benefits of psychoacoustic constraints and multi-scale processing.
\newblock {\em Connection Science}, 6(2-3):247--280, 1994.
\newblock \href {http://dx.doi.org/10.1080/09540099408915726}
  {\path{doi:10.1080/09540099408915726}}.

\bibitem{Pachet:2003wd}
François Pachet.
\newblock The continuator: Musical interaction with style.
\newblock {\em Journal of New Music Research}, 32(3):333--341, 2003.
\newblock \href {http://dx.doi.org/10.1076/jnmr.32.3.333.16861}
  {\path{doi:10.1076/jnmr.32.3.333.16861}}.

\bibitem{Sturm:2016rz}
Bob~L. Sturm, João~Felipe Santos, Oded Ben-Tal, and Iryna Korshunova.
\newblock Music transcription modelling and composition using deep learning.
\newblock In {\em Proceedings of the 1st Conference on Computer Simulation of
  Musical Creativity}, 2016.

\end{thebibliography}
\end{document}